\documentclass{article}
\usepackage{amssymb}
\usepackage{bm}
\usepackage[dvips]{graphicx}
\begin{document}
\title{Lower current-driven exchange switching threshold in noncollinear magnetic junctions under high spin injection}
\author{S.~G.~Chigarev, E.~M.~Epshtein\thanks{Corresponding author. E-mail: eme253@ms.ire.rssi.ru}, P.~E.~Zilberman\\ \\
V.~A.~Kotelnikov Institute of Radio Engineering and Electronics \\
Russian Academy of Sciences \\ Fryazino, Moscow District, 141190, Russia}
\date{}
\maketitle
\begin{abstract}
Current-induced switching is considered in a magnetic junction. The
junction includes pinned and free ferromagnetic layers which work in the
regime of the high spin injection. It is shown that in such a regime the
exchange magnetization reversal threshold can be lowered up to two times
when the axes of the layers are noncollinear.
\end{abstract}

\section{Introduction}\label{section1}
The effect of current-driven switching in magnetic junctions~\cite{Katine} attracts
continuing attention because of possible using this phenomenon for
high-density information processing. One of the most important problems
is lowering the current density threshold corresponding to instability of
the initial magnetic configuration and switching it to another one with
different resistance. Various ways have been proposed to solve the
problem, namely, using magnetic semiconductors with their lower saturation
magnetizations~\cite{Watanabe}, choosing the layer materials with proper spin
resistances~\cite{Gulyaev1}, and so on. In this note, we consider an additional
possibility of lowering the threshold current density corresponding to
instability and switching the magnetic configuration of the magnetic
junction.

\section{The model}\label{section2}
We consider a magnetic junction consisting of a pinned ferromagnetic layer
1, free ferromagnetic layer 2, an ultrathin spacer in between, and
nonmagnetic layer 3 closing the electric circuit. The current flows
perpendicular to the layer planes (CPP mode). The free layer thickness~$L$
is assumed to be small in comparison with the spin diffusion length in
that layer and the domain wall thickness. In such conditions, the
macrospin approximation~\cite{Gulyaev2} is applicable. In this approximation, the
modified Landau--Lifshitz--Gilbert (LLG) equation contains additional
terms describing two different mechanisms of the spin-polarized current
influence on the magnetic lattice, namely, the spin torque transfer
effect~\cite{Slonczewski,Berger} and the spin injection effect~\cite{Gulyaev3}. The relative role of these
mechanisms depends on the system parameters, specifically, the Gilbert
damping constant. It has been shown~\cite{Gulyaev2} that with suitable choice of the
layer spin resistances and not too small damping constant, the injection
mechanism predominates over the spin torque transfer mechanism. Such a
situation is assumed below.

We use a coordinate system with~$x$ axis along the current direction
and~$yz$ plane parallel to the layers plane. The easy axis of free layer 2
is parallel to~$z$ axis, while the easy axis of layer 1 is oriented at some
angle with respect to~$z$ axis.

\section{Switching conditions under high spin injection}\label{section3}
When the spin injection predominates over the spin torque transfer, the
LLG equation has the first integral in form of a magnetic energy
consisting of the Zeeman energy, the anisotropy energy, the
demagnetization energy, and the nonequilibrium \emph{sd} exchange energy
proportional to the spin-polarized current density~\cite{Elliott,Epshtein1}. The magnetic
energy is a function of the angles determining orientation of the free
layer magnetization with respect to the easy axis of the layer, the
external magnetic field, and the pinned layer magnetization (which is
parallel to the easy axis of this layer). The angles depend on the current
density because of the sd exchange interaction. The minima of the magnetic
energy correspond to the stable equilibrium states. The switching occurs
when the stable equilibrium state of the system disappears or converts to
unstable one (which corresponds to a maximum of the energy).

We assume below that the spin resistance~\cite{Epshtein2} of the free layer is much
lower than the spin resistances of the pinned and nonmagnetic layers.

Under assumptions mentioned, the (dimensionless) magnetic energy with the
current direction corresponding to the electron flow from the pinned layer
to the free one takes the form~\cite{Epshtein1}
\begin{equation}\label{1}
  U==\frac{H}{H_a}\cos(\theta-\beta)-\frac{1}{2}\cos^2\theta-\frac{j}{j_0}\cos(\theta-\theta_1),
\end{equation}
\begin{equation}\label{2}
  j_0=\frac{eH_aL}{\mu_{\rm B}\alpha\tau Q_1}.
\end{equation}
Here the following notations are used:~$H$ is the external magnetic
field,~$H_a$ is the anisotropy field of the free layer,~$\theta$ and~$\beta$ angles determine
directions of the free layer magnetization and the external magnetic
field, respectively, reckoned from the free layer easy axis ($z$ axis),~$\theta_1$
is analogous angle for the pinned layer,~$\alpha$ is the dimensionless constant
of \emph{sd} exchange interaction,~$\tau$ is the spin relaxation time,~$Q_1$ is the spin
polarization of the pinned layer conductivity,~$\mu_{\rm B}$ is the Bohr magneton,

In absence of the magnetic field ($H=0$) and the current ($j=0$), there are two
stable configurations,~$\theta=0$ and~$\theta=180^\circ$. At~$H=0$ and~$\theta_1=0$, the antiparallel relative
orientation of the ferromagnetic layers ($\theta=180^\circ$) becomes unstable, and
switching occurs to parallel configuration ($\theta=0$) at~$j=j_0$.

\section{Switching noncollinear configurations}\label{section4}
The stability analysis shows that the instability occurs at lower
threshold current density if the initial configuration is not collinear.
The minimal threshold current density~$j_{\rm{th}}=j_0/2$ takes place when the angle between
the easy axes of the pinned and free layers is equal to~$135^\circ$.
\begin{figure}
\includegraphics{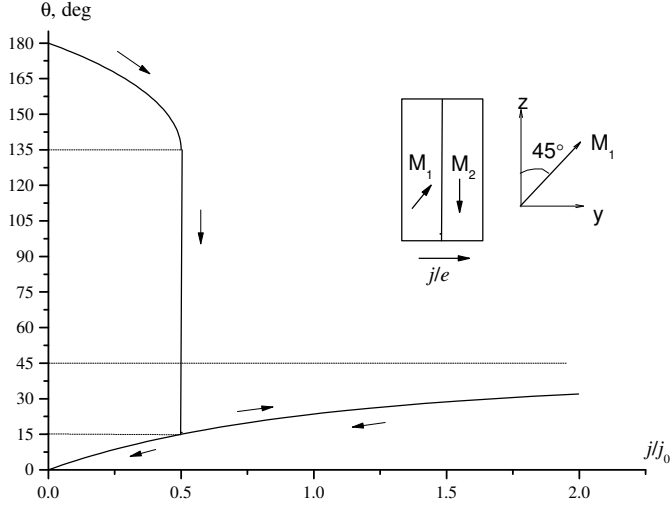}
\caption{Switching magnetic junction with noncollinear initial
configuration. The free layer magnetization orientation is shown versus
the current density through the junction. The arrows indicate direction
of the current change. Vectors~$\mathbf{M}_1$ and~$\mathbf{M}_2$ denote magnetizations of the pinned
and free layers, respectively.}\label{fig1}
\end{figure}

The numerical results are shown in fig.~\ref{fig1}. Let the pinned layer easy axis
and the magnetization vector~$\mathbf{M}_1$ be directed at the angle of~$\theta_1=45^\circ$ with respect
to~$z$ axis, while the initial (in absence of the current) direction of the
free layer magnetization corresponds to~$\theta=180^\circ$. When the current through
magnetic junction is turned on and increases, the magnetization vector~$\mathbf{M}_2$
of layer 2 deviates from the initial direction. When the current density
reaches value~$j=j_0/2$, that vector takes the position~$\theta=135^\circ$. At that moment,
instability occurs and vector~$\mathbf{M}_2$ abruptly turns to a new position~$\theta=15^\circ$. At
further increase in current, vector~$\mathbf{M}_2$ tends to stand parallel to~$\mathbf{M}_1$. If
the current is turns off, vector~$\mathbf{M}_2$ takes the position~$\theta=0$. Thus, the
switching of vector~$\mathbf{M}_2$ is realized from~$\theta=180^\circ$ to~$\theta=0$ position with threshold
current density~$j_{\rm{th}}=j_0/2$, half as much as the threshold for collinear
configuration.

If magnetic field~$H$ is applied along~$\mathbf{M}_1$ vector, the threshold current density reveals further decrease to
\begin{equation}\label{3}
    j_{\rm{th}}=\frac{j_0}{2}\left(1-\frac{2H}{H_a}\right).
\end{equation}
At~$H$ value close to but slightly less than the halved anisotropy field~$H_a/2$,
the instability threshold can be lowered considerably. Note, that
participation of the magnetic field does not prevent locality of the
effect, because the magnetic field~$H<H_a/2$ cannot do switching alone, without
the current.

The switching effect leads to change in resistance; this is of the same
origin as the well-known giant and tunnel magnetoresistance effects. The
current density through a tunnel junction with~$\chi=\theta-\theta_1$ angle between the layer
magnetizations is~\cite{Utsumi}
\begin{equation}\label{4}
  j=j_p\cos^2\frac{\chi}{2}+j_a\sin^2\frac{\chi}{2},
\end{equation}
where~$j_p$ and~$j_a$ are the current densities for parallel ($\chi=0$) and antiparallel
($\chi=180^\circ$) configurations, respectively. In the example considered ($\theta_1=45^\circ$),~$\chi$ angle
is equal to~$\chi_1=135^\circ-45^\circ=90^\circ$ just before the switching
and~$\chi_1=15^\circ-45^\circ=-30^\circ$ just after the switching.
Therefore, the junction resistance relative change due to the switching
is
\begin{equation}\label{5}
  \frac{R_1-R_2}{R_2}=\frac{\sqrt3}{2}\frac{\rho}{2+\rho},
\end{equation}
where~$\rho\equiv\left(j_p-j_a\right)/j_a$ is the standard magnetoresistance
ratio~\cite{Guilliere} corresponding to the
switching from one collinear configuration to another (opposite) one.

The threshold of the switching by a magnetic field can be lower up to two
times also when the field is applied at an angle to the easy axis of the
layer, in comparison with the collinear configuration~\cite{Chikazumi}. In our case,
the field is of exchange nature, but the result is similar. It should have
in mind, however, that the exchange switching, unlike the conventional
switching by a magnetic field, is localized at the atomic level. This fact
opens new practical possibilities.

\section{Conclusion}\label{section5}
The calculations show that the threshold current density corresponding to
switching the free layer magnetization to opposite direction can be
lowered up to two times if the pinned layer of the junction is
noncollinear with the easy axis of the free layer. Such an effect allows
varying conditions of experiments.

With more complex anisotropy when several easy axes present a many-level
switching is possible in a magnetic field~\cite{Uemura}. The generalization of this
result to the exchange switching by a spin-polarized current is of great
interest.

The work was supported by Russian Foundation for Basic Research, Grant
No.~08-07-00290.

\end{document}